\documentclass[epj]{svjour}
\usepackage{graphicx}
\newcommand{\NP}[1]{Nucl.\ Phys.\ {#1}}

\newcommand{\PL}[1]{Phys.\ Lett.\ {#1}}

\newcommand{\PR}[1]{Phys.\ Rev.\ {#1}}

\begin{document}
\title{Chiral condensate thermal evolution at finite baryon
chemical potential within ChPT}
\author{R. Garc\'\i{}a Mart\'\i{}n
\and J. R. Pel\'aez}
\institute{Dpto. de F\'\i{}sica Te\'orica II, UCM, 28040 Madrid, Spain.}
\abstract{
  We present a model independent 
  study of the chiral condensate evolution in a hadronic gas, in terms
  of temperature and baryon chemical potential. 
  The meson-meson interactions are described within Chiral Perturbation Theory 
  and the pion-nucleon interaction by means of Heavy Baryon Chiral Perturbation Theory,
  both at one loop. Together with the virial expansion, 
  this provides a model independent systematic expansion 
  at low temperatures and chemical potentials,
  which includes the physical quark masses. This
  can serve as a guideline for further studies on the lattice.
  We also obtain estimates of the
  critical line of temperature and chemical potential 
  where the chiral condensate melts, which systematically lie  somewhat higher
  than recent lattice calculations but are consistent with several hadronic models.
  \PACS{
    {11.30.Rd}{ } \and
    {11.30.Qc}{ } \and
    {12.38.Aw}{ } \and
    {12.39.Fe}{ } \and
    {11.10.Wx}{ }
  }
}
\date{Oct. 15, 2006}
\maketitle

On a recent work~\cite{paper} we approach the question of the
phase diagram of QCD. In particular, we study the transition from
the hadronic phase, in which the chiral symmetry is spontaneously
broken, to a phase in which it is restored. This is done within
Chiral Perturbation Theory (ChPT), the low energy effective
theory of QCD. We also gave estimates for the melting curve of the
quark condensate in presence of a baryon chemical potential.

Let us briefly describe ChPT. The spontaneous chiral symmetry breaking
of QCD requires the existence of eight massless Goldstone Bosons (GB),
that can be identified with the pions, kaons and eta. These are thus
the most relevant degrees of freedom at low energies. In addition, there
is an explicit symmetry breaking due to the non-vanishing quark masses
that give rise to a small mass for the pions, kaons and eta, which
are thus just pseudo-GB. ChPT is built as the most general derivative
and mass
expansion over the spontaneous symmetry breaking (SSB) scale
$\Lambda_\chi\equiv4\pi f_\pi\simeq1.2$~GeV
which respects the symmetry constraints of QCD. Baryons can be included
in this formalism, although its treatment is more involved due to their
large masses. Within Heavy Baryon ChPT (HBChPT)~\cite{Jenkins:1990jv}
this problem is overcome
for the meson-baryon interaction by an additional expansion on inverse
powers of the baryon mass.
This approach has proven very successful, and works remarkably well
within the meson sector, and with a somewhat slower convergence in the
meson-baryon sector.

Next, the thermodynamics of the hadron gas is described by the virial
expansion~\cite{virial,Gerber}. This is a low density
series expansion for the grand canonical potential of a relativistic
interacting multicomponent gas. We will assume that only the strong
interactions are relevant, and that the baryon density defined as
$n_B-n_{\bar B}$ is conserved. The thermodynamics of the gas will thus
depend on the temperature $T$ (usually written in terms of its
inverse $\beta=1/T$) and a baryon chemical potential $\mu_B$.
The virial expansion, written in terms of the pressure, reads:
\begin{equation}
\beta P = \sum_i B_i^{(1)}\,\xi_i + B_{i}^{(2)}\,\xi_i^2 +
\sum_i\sum_{j\ge i} B_{ij}^{int}\,\xi_i\xi_j + \dots,
\end{equation}
where $\xi_i=\exp\beta(\mu_i-M_i)$, with $M_i$ the mass of the
$i$-th species, $\mu_i=0,\pm\mu_B$ for mesons/(anti)baryons respectively.
The free virial coefficients
\begin{equation}
B_i^{(n)} = \frac{g_i \eta_i^{n+1}}{2\pi^2}\int_0^\infty dp\,p^2\,
e^{-n\beta(\sqrt{p^2+M_i^2}-M_i)},
\end{equation},
with $\eta_i=\pm1$ for bosons/mesons and $g_i$ the degeneracy of
the $i$-th species,
correspond simply to the series expansion of the pressure for a free
gas, while the interactions are taken into account in the term:
\begin{equation}
B_{ij}^{int}=\frac{e^{\beta(M_i+M_j)}}{2\,\pi^3}\int_{M_i+M_j}^{\infty} dE\, E^2 K_1(\beta E) 
\Delta^{ij}(E),
\end{equation}
where $K_1(x)$ is the modified Bessel function (of the second kind), and
$\Delta^{ij}=\sum_{I,J,S} (2I+1)(2J+1)\delta^{ij}$ are the
$ij\rightarrow ij$ elastic scattering phase shifts for a state
$ij$ with well defined isospin $I$, total angular momentum $J$
and strangeness $S$, defined so that $\delta=0$ at threshold.

The dependence of the chiral condensate with temperature and chemical
potential is written in terms of the pressure of the gas as follows:
\begin{eqnarray}
\langle\bar{q} q\rangle&=
&\langle 0 \vert\bar{q} q\vert 0 \rangle 
\left(1+\sum_h \frac{c_h}{2 M_h F^2} 
\frac{\partial P}{\partial M_h}\right),
\end{eqnarray}
where the constant $F$, which is the pion decay constant in the chiral
limit, is introduced for further convenience, and the coefficients
$
c_h=- F^2\frac{\partial M_h^2}
{\partial \hat m}\langle 0 
\vert\bar{q} q\vert 0 \rangle ^{-1}
\label{cis}
$
encode the hadron mass dependence on the quark mass. Numerically,
these coefficients amount to
$c_\pi = 0.9^{+0.2}_{-0.4}$,
$c_K = 0.5^{+0.4}_{-0.7}$,
$c_\eta = 0.4^{+0.5}_{-0.7}$,
for mesons~\cite{Pelaez:2002xf}, and
$c_N = 3.6^{+1.5}_{-1.9}$
for nucleons~\cite{paper}.

\begin{figure}
\begin{center}\includegraphics[scale=.9,angle=-90]{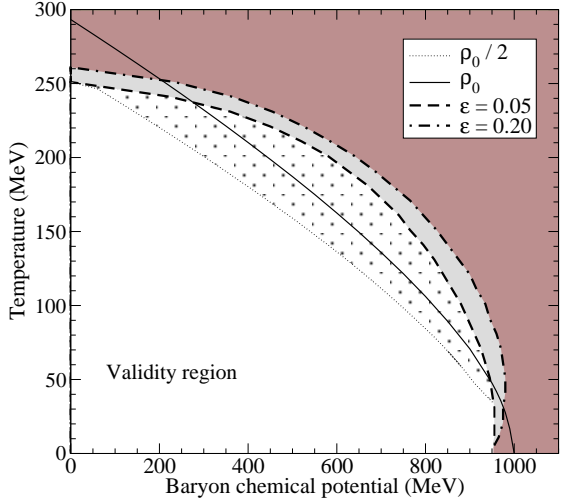}\end{center}
\caption{\label{fig:chkvir} \rm Crude estimate of the region
of validity of the virial
expansion for a free gas.
We have excluded the regions in which the virial expansion
breaks, but also those regions in which nuclear density is large
enough so that higher order effects are relevant.
}
\end{figure}

First of all, we want to estimate the region of validity of our
calculations. For this, we check the points for which the
virial expansion breaks down, by computing the difference between
the free condensate calculated exactly and with the virial expansion.
In Figure~\ref{fig:chkvir} we show, as a dashed and a dashed-dotted
lines, the points where the virial expansion is off the exact calculation 
by 5\% and 20\%, respectively. The expansion rapidly deteriorates after that.
Also, as we want to neglect $NN$ interactions, we have
to restrict ourselves to the region where nucleon density
is small enough. In ref.~\cite{Lutz} it is shown that, for densities
below $\rho_0/2$, with $\rho_0\simeq0.16\,{\rm fm}^{-3}$, and
{\it as far as we are only concerned with the quark condensate},
the $\pi N$ interaction dominates over the $NN$ interaction.
Thus, we restrict the validity of our calculation to the points
below the $\rho_0/2$ line. This line is shown in the Figure as a dotted
line, together with the $\rho_0$ line, which 
is shown for illustration as a continuous line. The white area in Fig.2 corresponds to our estimated ``validity region''.

\begin{figure}
\begin{center}\includegraphics[scale=.9,angle=-90]{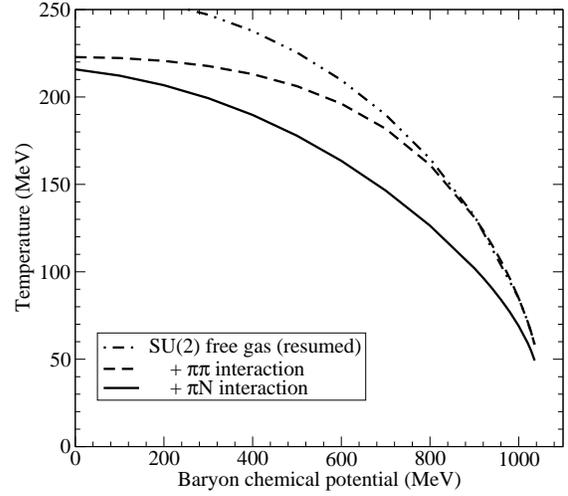}\end{center}
\caption{\label{fig:mut1} \rm 
Melting line in the $(\mu_B,T)$ plane of the chiral condensate
for an SU(2) gas of hadrons. We show the result for the free gas,
but also adding the $\pi\pi$ ChPT SU(2) interaction to one loop,
and the line resulting from adding the $\pi N$ interaction to third order
in HBChPT.
}
\end{figure}

In Figure~\ref{fig:mut1} we show in the $(\mu_B,T)$ plane
the condensate melting line of the free $SU(2)$ gas composed of
pions and nucleons as a dash-dot-dotted line. We have then added
the effects of the $\pi\pi$ interaction, which yields the dashed
line of the Figure. Finally, the $\pi N$ interaction was included.
We see that the effect of the $\pi N$ interaction is small, but
noticeable ($\sim 6$~MeV at $\mu_B=0$) at low baryon chemical potentials,
and maximum at around $\mu_B\simeq600-700$~MeV, where it produces
a decrease in the melting temperature of about $40$~MeV with respect
to the gas without $\pi N$ interactions.
Up to a chemical potential of $\mu_B=40-50$~MeV, the region of
relevance for Relativistic Heavy Ion Collisions, the decrease in the
melting temperature amounts roughly to $10$~MeV when we include
the $\pi N$ interaction.
It is important to note that, although the critical lines are actually
extrapolations, at low temperature and chemical potential our
calculation is model independent, and should be quite accurate.
Still, we show the melting lines since they 
 are useful to visualize and help quantifying the relative size of
each contribution we add into the gas.

In a real gas, we should consider all hadrons. We will do this by
including them as free particles. The only exception to this will
be the kaons and etas, which are abundant enough up to $200$~MeV
to deserve a separate treatment and include their interaction
with a pion. All other interactions are suppressed by
Boltzmann (thermal) and $c_h/M_h$ factors, and are therefore not
included in our treatment. At very large $\mu_B$ the heavier
nucleons may not be Boltzmann suppressed, and their interactions may
become important, but this is outside the scope of this work.
\begin{figure}
\begin{center}\includegraphics[scale=.9,angle=-90]{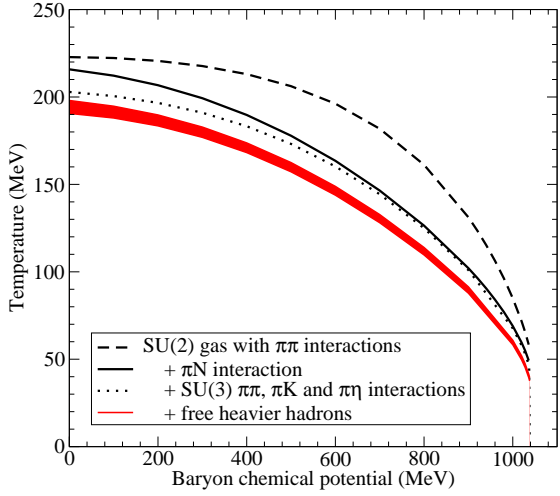}\end{center}
\caption{\label{fig:mut2} \rm Our final result for 
the chiral condensate melting line in the $(\mu_B,T)$ plane.
Starting from the SU(2) gas with $\pi\pi$ interactions,
we show the effect of $\pi N$ interactions and of adding
kaons and etas. The dark area covers the uncertainty due to 
heavier hadrons estimated as explained in the text.
}
\end{figure}
In Figure~\ref{fig:mut2} we show the two lines corresponding
to the $SU(2)$ gas with $\pi\pi$ interactions (dashed) and with
$\pi\pi$ and $\pi N$ interactions (continuous). We then plot
the line corresponding to a gas in which also kaons, etas,
and their interactions with a pion, now in $SU(3)$,
are included (dotted line).
Note that the effect is pronounced at $\mu_B=0$, with a decrease
of the melting temperature of around $\sim 13$~MeV, but becomes
negligible at very high baryon chemical potential.
Finally, we have added the effects of other heavier hadrons,
included as free particles, since  they suffer a larger Boltzmann suppression. We have estimated that, for heavier hadrons,
$\partial M_h / \partial \hat m
\simeq \alpha N_{u,d}^{h}$,
where $\alpha$ is an adimensional constant
and $N_{u,d}^{h}$ is the number
of valence $u$ or $d$ quarks in the hadron.
Based on the values of $\alpha$ that we have calculated for the
pions, kaons, etas and nucleons, we have estimated that,
for heavier hadrons, $\alpha\simeq0.5-2.5$. The band in
Figure~\ref{fig:mut2} corresponds to this range of values for $\alpha$.
In order to properly analyse the uncertainties in our calculation,
we also have to include the uncertainties that come from our imperfect
knowledge of the chiral parameters. This was done by adding in
quadrature the errors coming from independently varying each of the
parameters. The final results are plotted in Figure~\ref{fig:err},
where the dark band corresponds to the uncertainties coming from
the $\alpha$ parameter in heavy hadrons and the dashed line includes
also the uncertainty coming from the chiral parameters.
Note that the error is highly asymmetric:
even though every contribution lowers the melting temperature,
the melting of the condensate accelerates near the melting point,
and thus even if the contribution has a similar size to the previous
one, its effect on the condensate melting seems smaller.
\begin{figure}
\begin{center}\includegraphics[scale=.9,angle=-90]{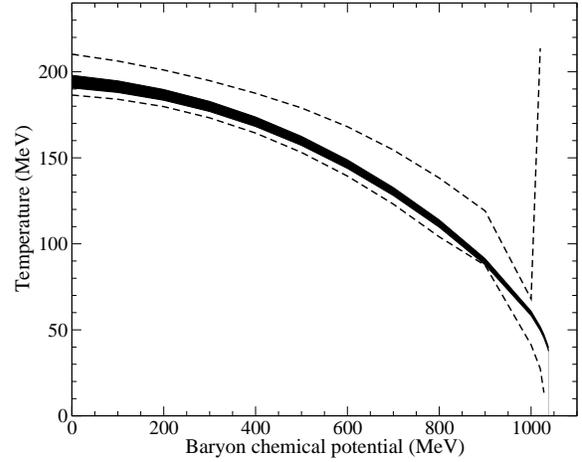}\end{center}
\caption{\label{fig:err} 
\rm Analysis of uncertainties. To the dark area, which corresponds
to the case described in Fig.~\ref{fig:mut2}, we have added
the uncertainties coming from the chiral parameters, shown as
dashed lines (see~\cite{paper} for discussion).
}
\end{figure}

\begin{figure}
\begin{center}\includegraphics[scale=.9,angle=-90]{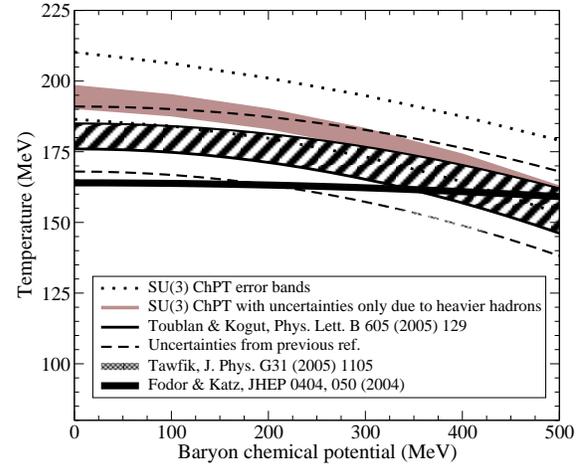}\end{center}
\caption{Comparison between our results and others in the literature.
Our melting temperatures are consistent within errors with
ref.~\cite{Toublan:2004ks} but systematically higher than
\cite{lattice}.}
\label{fig:lat}
\end{figure}

It is interesting to compare our results with those obtained by using
lattice calculations.
In Figure~\ref{fig:lat} we show our curves with
their errors, plotted against other results recently appeared in the
literature. Our approach agrees, within errors, with the hadron
resonance gas model described in~\cite{Toublan:2004ks}.
However, 
ChPT calculations \cite{Gerber,Pelaez:2002xf}, including ours \cite{paper}, yield melting temperatures that are
systematically above the results from the lattice
calculations~\cite{lattice}.


\end{document}